\documentclass[twocolumn]{jpsj3} 
\def\vec#1{\mbox{\boldmath $#1$}}

\title{$^{73}$Ge-NMR/NQR Investigation of Magnetic Properties of URhGe}

\author{Hisashi \textsc{Kotegawa}$^{1}$\thanks{E-mail address: kotegawa@crystal.kobe-u.ac.jp}, Kenta \textsc{Fukumoto}$^{1}$, Toshihiro \textsc{Toyama}$^{1}$, Hideki \textsc{Tou}$^{1}$, \\
Hisatomo \textsc{Harima}$^{1}$, Atsushi \textsc{Harada}$^{2}$, Yoshio \textsc{Kitaoka}$^{2}$, Yoshinori \textsc{Haga}$^{3}$, \\
Etsuji \textsc{Yamamoto}$^{3}$, Yoshichika \textsc{\=Onuki}$^{3,4}$, Kohei M. \textsc{Itoh}$^{5}$ and Eugene E. \textsc{Haller}$^{6}$}

\inst{$^1$Department of Physics, Kobe University, Kobe 657-8501, Japan\\
$^2$1Graduate School of Engineering Science, Osaka University, Osaka 560-8531, Japan\\
$^3$ASRC, JAEA, Tokai, Ibaraki 319-1195, Japan\\
$^4$Faculty of Science, University of the Ryukyus, Nishihara, Okinawa 903-0213, Japan\\
$^5$School of Fundamental Science and Technology, Keio University, Kohoku-ku, Yokohama 223-8522, Japan\\
$^6$Department of Materials Science and Engineering, University of California at Berkeley and Lawrence Berkeley National Laboratory, Berkeley, California 94720, USA

}

\abst{
We report $^{73}$Ge-NMR and NQR results for ferromagnetic (FM) superconductor URhGe.
The magnitude and direction of the internal field, $H_{\rm int}$, and parameters of the electric field gradient at the Ge site were determined experimentally.
Using powdered polycrystalline samples oriented by different methods, the field dependences of NMR shift and nuclear spin relaxation rates for $H_0 \parallel c$ (easy axis) and $H_0 \parallel b$ were obtained.
From the NMR shifts for $H_0 \parallel b$, we confirmed a gradual suppression of the Curie temperature and observed a phase separation near the spin reorientation.
The observation of the phase separation gives microscopic evidence that the spin reorientation under $H_0 \parallel b$ is of first order at low temperatures.
The nuclear spin-lattice relaxation rate $1/T_1$ indicates that the magnetic fluctuations are suppressed for $H_0 \parallel c$, whereas the fluctuations remain strongly for $H_0 \parallel b$.
The enhancements of both $1/T_1T$ and the nuclear spin-spin relaxation rate $1/T_2$ for $H_0 \parallel b$ toward the spin reorientation field suggest that the field-induced superconductivity in URhGe emerges under the magnetic fluctuations along the $b$ axis and the $c$ axis.
}

\kword{URhGe, spin reorientation, magnetic fluctuation, NMR}

\begin{document}
\maketitle

\section{Introduction}

Intimate interplay between superconductivity and magnetism has been intensively investigated in many unconventional superconductors such as cuprates, heavy fermion systems, organic superconductors, and Fe-pnictides.
Superconductivity is realized in various situations; however, among them, novel superconductivity coexisting with ferromagnetism emerges only in uranium compounds such as UGe$_2$, URhGe, UCoGe, and UIr.\cite{Saxena,DAoki,Huy,Akazawa}
The first three compounds have several similarities; for instance, they have crystarographically similar U-zigzag chains, and each magnetic anisotropy is an Ising-type.
On the other hand, quantitative differences appear in some aspects.
Superconductivity in URhGe occurs at $T_{sc}=0.25$ K well below a Curie temperature of $T_{\rm Curie}=9.5$ K.\cite{DAoki}
The ordered moment of $0.42\mu_B/$U lies along the $c$ axis.\cite{DAoki}
These values are different among three compounds; $T_{sc}=0.6$ K, $T_{\rm Curie}=2.8$ K, and $0.03\mu_B/$U for UCoGe,\cite{Huy} and $T_{sc}=0.7$ K, $T_{\rm Curie}\sim30$ K, and $0.9\mu_B/$U for UGe$_2$ (at 1.2 GPa).\cite{Saxena}
A comparison among these compounds is essential to understand what is an important factor in the ferromagnetic (FM) superconductors.

Another notable phenomenon is a field-induced superconductivity, which is seen remarkably in URhGe when the magnetic field was applied along the $b$ axis.\cite{Levy}
Superconductivity disappears once by applying the magnetic field of more than 2 T, but the system reenters the superconducting (SC) phase above 8 T.
The torque measurements have suggested that the spin reorientation from the $c$ axis to the $b$ axis occurs at the spin reorientation field, $H_R\sim12$ T.
In UCoGe and UGe$_2$, SC phases are not separated, but unusual field dependences of the upper critical field $H_{c2}$ are seen.\cite{Aoki_UCoGe,Huxley_PRB}

Nuclear magnetic resonance (NMR) and nuclear quadrupole resonance (NQR) measurements have contributed to promote an understanding of the FM superconductors.
In UGe$_2$ and UCoGe, the microscopic coexistence of superconductivity and ferromagnetism, microscopic states of the phase boundary, a character of the magnetic fluctuations, and spin susceptibility have been confirmed and discussed.\cite{Kotegawa_UGe2,Harada,Harada_PRB,Ohta,Ihara,Hattori,Hattori_PRB,Hattori_JPSJ}
However, no NMR/NQR experimental data have been reported on URhGe.
In this paper, we report $^{73}$Ge-NMR/NQR results for URhGe to reveal the magnetic properties. URhGe belongs to $P_{nma}$ space group possessing one Ge site, and the local symmetry at the Ge site is expressed by [$.m.$].

\section{Experimental Procedure}

Polycrystalline sample was prepared using enriched $^{73}$Ge.
Unfortunately a quality of the sample is not good enough to induce superconductivity; residual resistivity ratio (RRR) was about 4.
In this paper, we focus the magnetic properties of URhGe, which are not considered to be so sensitive to the sample quality, because the spin reorientation has been observed at the similar magnetic fields for the samples, whose RRR values range from 12 to 40.\cite{Miyake}
NQR measurement was performed using the powdered sample, and the sample was oriented both along $H_0 \parallel c$ and $H_0 \parallel b$ for NMR measurements.
The nuclear spin-lattice relaxation time $T_1$ was measured at the $\pm7/2 \leftrightarrow \pm9/2$ transition for the paramagnetic (PM) state, the $+1/2 \leftrightarrow +3/2$ transition for the FM state at zero field, and at the central ($-1/2 \leftrightarrow +1/2$) transition for the NMR measurements.
Each recovery curve was fitted to the theoretical function corresponding to the transition to estimate $T_1$.
The nuclear spin-spin relaxation time $T_2$ was measured at the central transition for $H_0 \parallel b$, and the decay curves were fitted to the single exponential function.

\section{Experimental Data}

\subsection{NQR/NMR spectral analysis}

Figure 1 shows $^{73}$Ge-NQR spectra measured at zero field.
In the PM phase at 15 K, three peaks are observed in the frequency range from 2 to 4.5 MHz.
Each peak corresponds to $\pm3/2 \leftrightarrow \pm5/2$ ($\sim2.1$ MHz), $\pm5/2 \leftrightarrow \pm7/2$ ($\sim3.2$ MHz), and $\pm7/2 \leftrightarrow \pm9/2$ ($\sim4.2$ MHz) transition.
These resonance frequencies provide the quadrupole frequency $\nu_Q=1.06$ MHz and the asymmetry parameter $\eta=0.09$.
In the ordered state at 1.4 K, the emergence of the internal field modifies the spectral shape.
The spectrum is complicated, but it can be almost reproduced by simulations using the following parameters; (red curve: $H_{\rm int}=1.70$ T, $\theta=53^{\circ}$, $\phi=90^{\circ}$, $\nu_Q=1.03$ MHz and $\eta=0.08$), or (blue curve: $H_{\rm int}=1.72$ T, $\theta=53^{\circ}$, $\phi=90^{\circ}$, $\nu_Q=0.96$ MHz and $\eta=0.07$).
Here, $H_{\rm int}$ is a magnitude of the internal field at the Ge site, and respective $\theta$ and $\phi$ are angles between the principal axes of the electric field gradient (EFG) and the effective field at the Ge site, $H_{\rm eff}$, which is equal to $H_{\rm int}$ under zero external field.
It was difficult to determine the better set of parameters between two sets of parameters.
The red curve does not reproduce the signal at around 5 MHz, but the $\nu_Q$ and $\eta$ are consistent with those obtained from the NMR spectrum.

\begin{figure}[htb]
\centering
\includegraphics[width=0.65\linewidth]{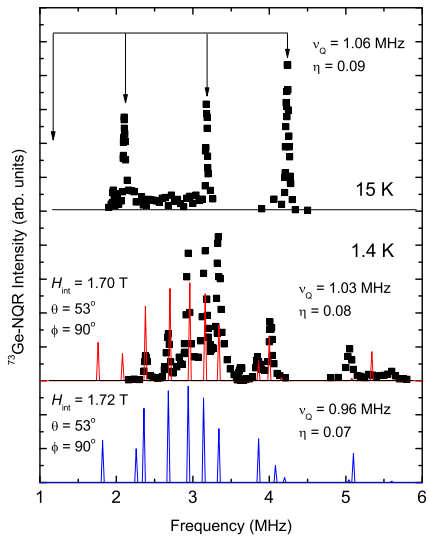}
\caption[]{(color online) $^{73}$Ge-NQR spectra measured at zero field in the PM and FM states. In the PM state, three transitions were observed among four transitions for $I=9/2$. The complicated spectrum in the FM state can be almost reproduced using the parameters described in the figure. We show two sets of parameters, but it was difficult to determine the better set of parameters.
}
\end{figure}

\begin{figure}[htb]
\centering
\includegraphics[width=0.9\linewidth]{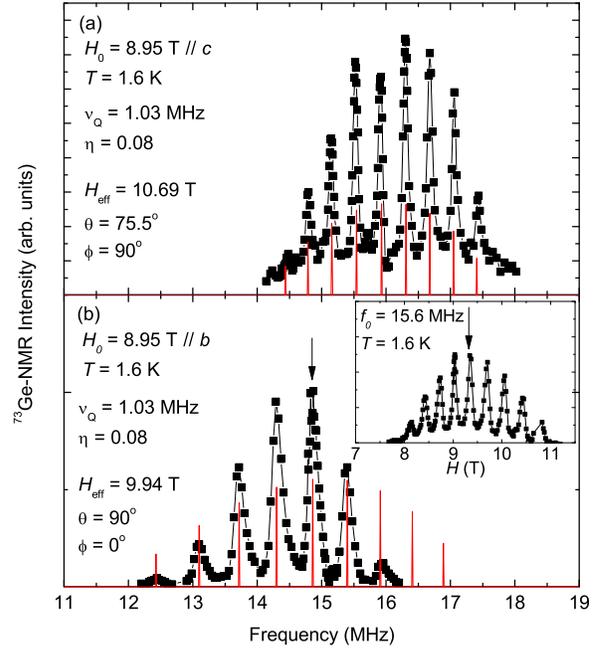}
\caption[]{(color online) $^{73}$Ge-NMR spectra of URhGe for the oriented samples along (a) $H_0 \parallel c$ and (b) $H_0 \parallel b$. At 8.95 T, each spectrum was observed in the different frequency ranges. The inset shows field-swept spectrum with fixed frequency for $H_0 \parallel b$. The central transition is observed at $\sim9.4$ T, which corresponds to the peak at $\sim14.8$ MHz in the frequency-swept spectrum as shown by arrows. In the frequency-swept spectrum, the measurement above 16.2 MHz has not been performed. The sharp peaks in all the spectra ensure the high degree of the orientation. 
}
\end{figure}

Figures 2 (a) and (b) show $^{73}$Ge-NMR spectra measured under external magnetic fields.
The spectrum in Fig.~2(a) was obtained after giving a mechanical shock to the powdered sample at low temperatures so as to orient the magnetically easy axis of the crystal, that is, the $c$ axis toward the magnetic field.
The spectrum consists of the well-separated nine peaks, which ensure a high degree of the orientation.
The spectrum can be reproduced by the parameters of $H_{\rm eff}=10.69$ T, $\theta=75.5^{\circ}$, $\phi=90^{\circ}$, $\nu_Q=1.03$ MHz and $\eta=0.08$.
Here, $H_{\rm eff}$ is an effective field at the Ge site, which is a vector sum of the external field $H_0$ and $H_{\rm int}$, that is, $\vec{H_{\rm eff}} = \vec{H_0} + \vec{H_{\rm int}}$.

A first principle calculation gave that the direction of the first principal axis of the EFG, $V_{zz}$ is tilted somewhat from the $a$ axis to the $c$ axis, and the second principal axis lies along the $b$ axis;\cite{Vzz} thus, we adopted that $V_{zz}$ ($\theta=0^{\circ}$) lies in the $ac$ plane and the second principal axis ($\theta=90^{\circ}$, $\phi=0^{\circ}$) lies along the $b$ axis as shown in Fig.~3.
The angles of $\phi=90$$^{\circ}$ obtained in Fig.~1 and Fig.~2(a) indicate that respective $H_{\rm int}$ and $H_{\rm eff}$ are in the $ac$ plane.
From the zero-field spectrum, the angle between $H_{\rm int}$ and $V_{zz}$ is deduced experimentally as 53$^{\circ}$, and the angle between $H_{\rm eff}$ and $V_{zz}$ is determined as 75.5$^{\circ}$ from the NMR spectrum.
Next, we assumed that the direction of $H_{\rm int}$ is independent of the external field applied along the $c$ axis.
This is consistent with an assumption that the hyperfine coupling tensor is independent of $H_0$, and the magnetic moment is directed to the $c$ axis both at zero field and under $H_0 \parallel c$.
Then, the angle between $H_{\rm eff}$ and $H_{\rm int}$ is estimated as $75.5^{\circ}-53^{\circ}=22.5^{\circ}$.
Using this angle and the relation of $\vec{H_{\rm eff}} = \vec{H_0} + \vec{H_{\rm int}}$, the internal field $H_{\rm int}=1.92$ T under $H_0=8.95$ T can be estimated.
Similarly, the angle between $H_{\rm int}$ and $H_0$, which is along the $c$ axis, can be determined as 27.2$^{\circ}$.
Consequently, we evaluated that $V_{zz}$ is tilted by 9.8$^{\circ}$ from the $a$ axis, as shown in Fig.~3.

\begin{figure}[htb]
\centering
\includegraphics[width=0.8\linewidth]{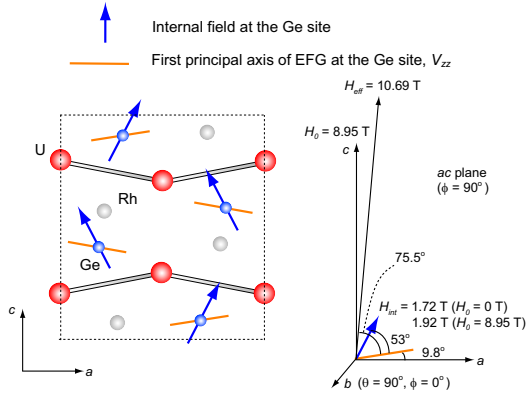}
\caption[]{(color online) The relationship between the crystal axes, the principal axes of the EFG, and each magnetic field, $H_{\rm int}$, and $H_{\rm eff}$. It is assumed that $V_{zz}$ ($\theta=0^{\circ}$) lies in the $ac$ plane and the second principal axis ($\theta=90^{\circ}$, $\phi=0^{\circ}$) lies along the $b$ axis from the first principle calculation. The $H_{\rm int}$ at the Ge site is tilted slightly from the $c$ axis. This is expected to originate from the anisotropy of the hyperfine coupling constants. 
}
\end{figure}

On the other hand, the spectra in Fig.~2(b) and its inset were obtained by a different method from Fig.~2(a). 
The sample was fixed in the paraffin at $\sim60$ $^{\circ}$C under 10 T; thus, the crystals were oriented along the easy axis at high temperatures, which was the $b$-axis.
As shown in Fig.~2(a) and 2(b), the two spectra measured in the same magnetic field and temperature are observed in the different frequency ranges.
The inset shows the field-swept spectrum in which a central transition was observed at $\sim9.4$ T.
In the frequency-swept spectrum, this transition corresponds to the peak at $\sim14.8$ MHz, which is lower than the central transition of $\sim15.9$ MHz for $H_0 \parallel c$, indicating that the internal field along $H_0$ is much smaller than the case of $H_0 \parallel c$.
The spectrum in Fig.~2(b) is reproduced by $\theta=90$$^{\circ}$ and $\phi=0$$^{\circ}$, which are the angles obtained when $H_{\rm eff}$ is parallel to the $b$ axis.
Actually, the directions of $H_0$ and $H_{\rm eff}$ are slightly different owing to the presence of $H_{\rm int}$, but this contribution is expected to be small because of $H_0 >> H_{\rm int}$.
Therefore the simulation indicates that the crystals are oriented along $H_0 \parallel b$.
The orientation along $H_0 \parallel b$ is also confirmed by a signature of the field-induced spin reorientation, as described later.

\subsection{Temperature dependences of the NMR shifts}

Figure 4 (a) shows the temperature dependence of NMR shift defined as $H_{\rm eff} -H_0$ for $H_0 \parallel c$ (easy axis).
The increase of the shift at $\sim10-20$ K originates in the development of the magnetization along the $c$ axis, $M_c$.
The shifts at 1.6 K under 3, 5, and $\sim9$ T and the magnetization data \cite{Hardy} gave a hyperfine coupling constant along the $c$ axis as $A^c \sim 3.24$ T/$\mu_B$.
Here, $A^c$ is defined as $H_{\rm eff} -H_0 = A^c M_c$.
The temperature derivatives of the shifts show peaks, which depend on $H_0$ as shown in Fig.~4(b).
This temperature corresponding to the increase of the magnetization is estimated to be $\sim14$ K at $H_0=3$ T and $\sim18$ K at $H_0=5$ T.
These are apparently higher than $T_{\rm Curie}=9.5$ K at zero field, indicating the transition is a crossover under the magnetic field along the easy axis.

\begin{figure}[htb]
\centering
\includegraphics[width=0.75\linewidth]{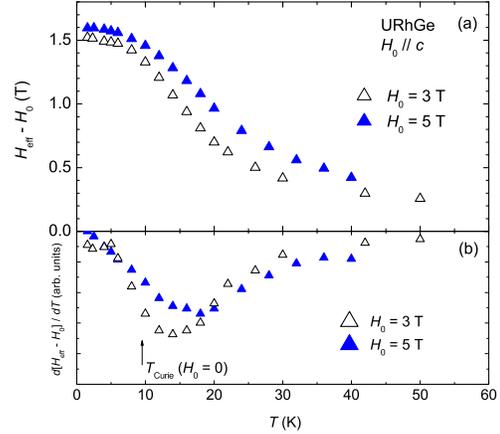}
\caption[]{(color online) The temperature dependence of NMR shift $H_{\rm eff} -H_0$ for $H_0 \parallel c$. The shift increases toward the low temperatures because of a development of the static moment. Their derivatives have a peak at the characteristic temperature of a crossover, which is higher than the original $T_{\rm Curie}$.}
\end{figure}

\begin{figure}[htb]
\centering
\includegraphics[width=0.65\linewidth]{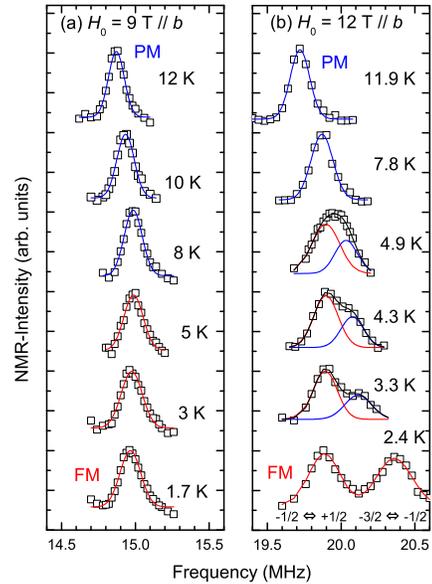}
\caption[]{(color online) NMR spectra of the central transitions for $H_0 \parallel b$. At 9 T, the peak position shifts to the higher frequency down to $\sim8$ K, and then returns to the lower frequency slightly. This temperature of $\sim8$ K corresponds to $T_{\rm Curie}$ under $H_0 \parallel b$. At 12 T, the spectrum is composed of two components at around $T_{\rm Curie}$, indicating that the transition is of a first order accompanied by the phase separation.
}
\end{figure}

Figure 5(a) and (b) show the NMR spectra for the central transition at several temperatures for $H_0 \parallel b$.
At $H_0 = 9$ T, the peak position shifts to the higher frequency with decreasing temperature down to $\sim8$ K, and then returns to the lower frequency slightly.
All the spectra are reproduced by a single Gaussian function.
The temperature dependences of NMR shift $H_{\rm eff} -H_0$ estimated from the resonance frequencies are shown in Fig.~6(a).
The shift at $H_0 = 9$ T increases with decreasing temperature down to $\sim7-8$ K, at which it has a broad maximum.
This temperature corresponds to $T_{\rm Curie}$ under $H_0 \parallel b$, which decreases from the original value of $T_{\rm Curie}=9.5$ K.\cite{Hardy,Miyake}
The shifts above 9 T show that $T_{\rm Curie}$ decreases with further application of the magnetic field.
On the other hand, the hyperfine coupling constant along the $b$ axis was estimated to be $A^b \sim 3.25$ T/$\mu_B$ from a comparison between the shift and the magnetization data in the PM state at 9 T,\cite{Hardy} 
This value is almost same as $A^c=3.24$ T/$\mu_B$.

As shown in Fig.~5(b), at $H_0 = 12$ T, the behavior is similar to the case of 9 T at high temperatures, and the spectrum is reproduced by a single Gaussian function, 
However, the spectrum becomes broader with decreasing temperature.
Especially the spectra at 4.3 K and 3.3 K are composed of two components, and each spectrum can be reproduced by a summation of two Gaussian functions as shown by red and blue curves.
The overlap of two Gaussian functions gives a microscopic evidence of the phase separation on a discontinuous phase transition.
This is not attributed to extrinsic origins such as a misalignment of the crystals.
With decreasing temperature, the component at the higher frequency shifts to the high-frequency side accompanied by a decrease in the intensity, and seems to overlap at the first satellite peak ($-3/2 \leftrightarrow -1/2$ transition) at 2.4 K.
We could not observe any signature of the phase separation below 2.4 K, and judged that the system is in the homogeneous FM state.
The NMR shift at 12 T is shown in Fig.~6(a), where two data are plotted at the same temperature because of the phase separation.
The shift corresponding the PM phase continues to increase toward low temperature.
This tendency is consistent with the magnetization data above 12 T.\cite{Hardy}
The $T_{\rm Curie}$ was estimated to be $4\pm1$ K from the NMR shift at 12 T.

Figure 6(b) shows the temperature dependence of the spectral line width under each magnetic field.
Here, the line width was estimated from the full width at the half maximum even for the asymmetric spectra at the phase separation region.
The temperature dependences of the line width below 10 T are monotonous, suggesting that the FM transition is of second order below 10 T.
At 12 T, apparent increase of the line width is observed, and this tendency remains at 11 and 11.5 T.
This may originate in the phase separation, but the broad spectra at 11 and 11.5 T are almost reproduced by a single Gaussian function.
Therefore, we cannot exclude a possibility that the broadening at 11 and 11.5 T originates in the misalignment of the oriented crystals.


\begin{figure}[tb]
\centering
\includegraphics[width=\linewidth]{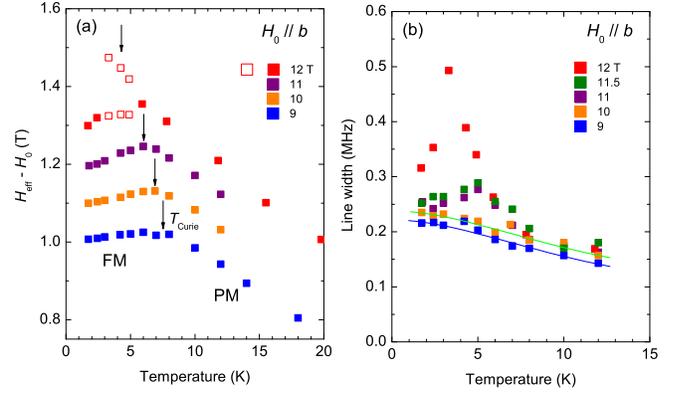}
\caption[]{(color online) Temperature dependences of (a) NMR shift and (b) line width for $H_0 \parallel b$. The $T_{\rm Curie}$ is determined at the peak position of the shift. At 12 T, the clear phase separation is observed in the temperature range shown by the open symbols. The shift in the PM state continues to increase toward low temperature. 
}
\end{figure}

\begin{figure}[htb]
\centering
\includegraphics[width=0.7\linewidth]{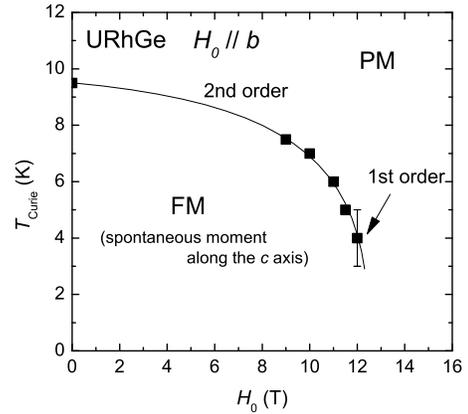}
\caption[]{(color online) The magnetic field-temperature phase diagram of URhGe for $H_0 \parallel b$. The TCP of URhGe is expected to be located between (10 T, 7 K) and (12 T, 4 K). In the present experimental setting, $H_R$ is roughly estimated to be $\sim13$ T.
}
\end{figure}

The magnetic field - temperature phase diagram for $H_0 \parallel b$ is shown in Fig.~7, where $T_{\rm Curie}$ is determined by the NMR shift.
We distinguished that the transition below 10 T is of second order, and that at 12 T is of first order accompanied by the phase separation.
From our experiments, the tricritical point (TCP), at which the second-order transition changes to first-order, is roughly estimated to be located between (10 T, 7 K) and (12 T, 4 K).
The presence of the TCP is consistent with other experimental and theoretical indications.\cite{Levy,Yelland,Aoki_Hall,Mineev}
The rough extrapolation of $T_{\rm Curie}$ indicates that the spin reorientation field at 0 K is present at $H_R\sim13$ T in this experimental setting.
The torque measurement using a single crystal has shown that a misalignment of 2.5$^{\circ}$ from the $b$ axis to the $c$ axis induces the increase in $H_R$ from $\sim12$ T to $\sim13$ T.\cite{Levy}
Although a direct comparison between the single crystal and the powdered polycrystalline sample is difficult, the average misalignment in this setting is expected to be $2-3^{\circ}$.

\subsection{Temperature dependences of nuclear spin-lattice relaxation}

\begin{figure}[htb]
\centering
\includegraphics[width=0.65\linewidth]{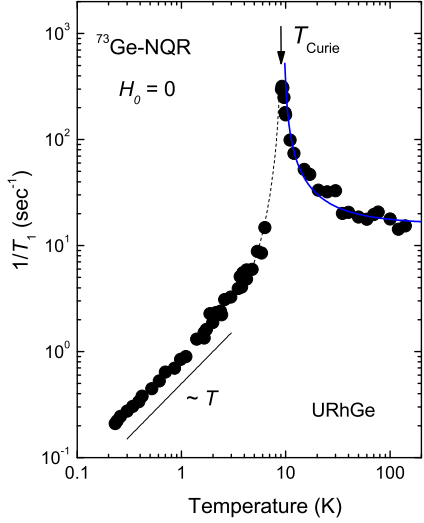}
\caption[]{(color online) Temperature dependence of $1/T_1$ measured at zero field for URhGe. The blue curve shows $1/T_1 \propto T/(T-T_{\rm Curie})$ expected in three dimensional itinerant ferromagnet.
}
\end{figure}

Figure 8 shows the temperature dependence of $1/T_1$ measured at zero field for URhGe.
With decreasing temperature, $1/T_1$ shows a divergent behavior toward $T_{\rm Curie}$, followed by a gradual decrease below $T_{\rm Curie}$ because of a gap-opening in the magnetic excitation.
$1/T_1$ well below $T_{\rm Curie}$ shows a $T_1T=const.$ behavior, which is same as the cases in the FM1 phase of UGe$_2$ (for example, 1.2 GPa) and UCoGe.\cite{Kotegawa_UGe2,Ohta}
In the PM state for URhGe, $1/T_1$ obeys $1/T_1 \propto T/(T-T_{\rm Curie})$, which is expected in three dimensional itinerant ferromagnet.\cite{Moriya}
The itinerant character of $f$ electrons has been confirmed by angle-resolved photoelectron spectroscopy, although the measurements have been limited below 20 K.\cite{Fujimori}
The behavior in $1/T_1$ in the PM state differs from that in UCoGe,\cite{Ohta} where $1/T_1$ shows a gradual decrease below $T^* \sim 40$ K.
This $T^*$ has been interpreted as a characteristic temperature, below which the itinerant character of the $5f$ electrons is enhanced.
It is conjectured that this difference between the materials originates from the difference in the band structures, and the similar difference is seen in the susceptibility measurements, where the susceptibility maximum is seen at $\sim40$ K for UCoGe, whereas such behavior is not seen in URhGe.\cite{Knafo}

\begin{figure}[htb]
\centering
\includegraphics[width=0.8\linewidth]{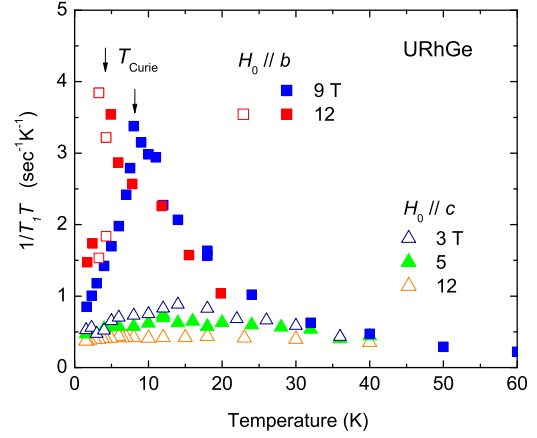}
\caption[]{(color online) Temperature dependences of $1/T_1T$ under the different magnetic fields, $H_0 \parallel c$ and $H_0 \parallel b$. For $H_0 \parallel c$, $1/T_1$ shows the broad peak at the crossover temperature. In contrast, $1/T_1T$ for $H_0 \parallel b$ shows sharp peaks at $T_{\rm Curie}$. At $H_0 = 12$ T, $1/T_1T$'s shown by the open symbols indicate those measured at the phase-separated PM and FM sates.
}
\end{figure}

\begin{figure}[b]
\centering
\includegraphics[width=0.7\linewidth]{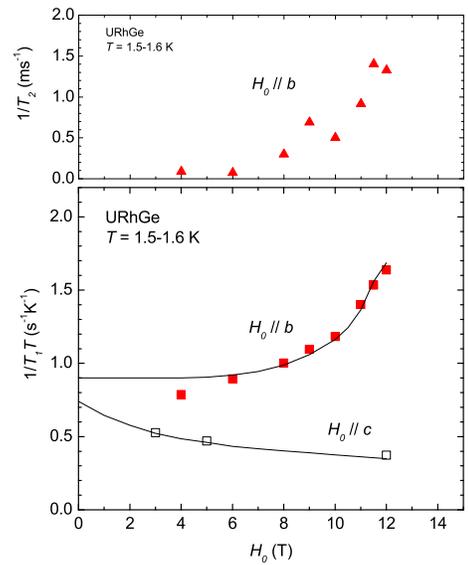}
\caption[]{(color online) Temperature dependences of $1/T_1T$ and $1/T_2$. $1/T_1T$ for $H_0 \parallel b$ corresponds to the magnetic fluctuations along the $ac$ plane and $1/T_2$ for $H_0 \parallel b$ corresponds to the magnetic fluctuations along the $b$ axis. The increases in both $1/T_1T$ and $1/T_2$ toward $H_R$ indicate that the magnetic fluctuations along the $b$ and $c$ axes are enhanced at around the spin reorientation. The curves indicate $c_i \gamma_S^2(H)$ (see text).
}
\end{figure}

Figure 9 shows the temperature dependence of $1/T_1T$ under the different magnetic fields.
For $H_0 \parallel c$, $1/T_1T$ has a broad maximum at $\sim14$ K at $H_0 = 3$ T, and the peak is smeared out under higher magnetic fields.
The peak $\sim14$ K at $H_0 = 3$ T is in good agreement with the crossover temperature, below which the NMR shift increases steeply as shown in Fig.~4.
The broadened anomaly in $1/T_1T$ suggests that the magnetic fluctuations are suppressed by the magnetic field along the easy axis, even for 3 T.
In the case of $H_0 \parallel b$, on the other hand, the magnetic fluctuations remain strongly under the magnetic field.
$1/T_1T$ at 9 T shows a sharp peak at $\sim8$ K, which is consistent with $T_{\rm Curie} \sim 7-8$ K estimated by the NMR shift shown in Fig.~6.
At 12 T, $1/T_1T$ continues to increase in the PM state, and it is suppressed in the FM state through the phase-separation region.

\subsection{Field dependences of nuclear spin-lattice relaxation and nuclear spin-spin relaxation}

Figure 10 shows the field dependences of $1/T_1T$ and $1/T_2$ at low temperatures ($1.5-1.6$ K).
$1/T_1T$ for $H_0 \parallel c$ decreases with increasing field, whereas $1/T_1T$ for $H_0 \parallel b$ shows a marked increase toward 12 T, which is close to the spin-reorientation field $H_{R}$. 
The increase of $1/T_1T$ under $H_0 \parallel b$ is qualitatively consistent with the behavior in UCoGe.\cite{Hattori_JPSJ}
In general, $1/T_1$ is expressed as follows.
\begin{eqnarray}
\frac{1}{T_1} = \frac{\gamma_N^2}{2} \int_{-\infty}^{\infty} \langle \delta H^-(t) \delta H^+(0) \rangle \exp(-i \omega_N t) dt
\end{eqnarray}
Here $\langle \delta H^-(t) \delta H^+(0) \rangle$ is time correlation function for magnetic fluctuations perpendicular to the quantum axis of the nuclear spin, and $\omega_N$ is the resonance frequency.
Thus, $1/T_1$ corresponds to the magnetic fluctuations perpendicular to the external magnetic field.
Another expression for $1/T_1T$ measured under the magnetic field along $\alpha$ axis is as follows.
\begin{eqnarray}
\left(\frac{1}{T_1T} \right)_{\alpha} = \frac{\gamma_N^2 k_B}{\gamma_e^2 \hbar^2} \sum_q \left[ A_{\beta}^2 \frac{\chi"_{\beta}(q,\omega_N)}{\omega_N} + A_{\gamma}^2 \frac{\chi"_{\gamma}(q,\omega_N)}{\omega_N} \right]
\end{eqnarray}
Here, $\chi"_{\beta}(q,\omega_N)$ [$\chi"_{\gamma}(q,\omega_N)$] are imaginary parts of dynamical susceptibility along $\beta$ [$\gamma$] axes, which are at right angles to $\alpha$.
The difference in $1/T_1T$ between $H_0 \parallel b$ and $H_0 \parallel c$ is explained by the anisotropy of the magnetic fluctuations as discussed in UCoGe;\cite{Ihara,Hattori} however, we should discuss anisotropy of the magnetic fluctuations using the extrapolation values toward $H_0 \rightarrow 0$, because the field dependences of $1/T_1T$ are present oppositely between $H_0 \parallel b$ and $H_0 \parallel c$.

To know the extrapolation values toward $H_0 \rightarrow 0$, we show referential curves represented by $c_i \gamma_S^2(H)$, where $\gamma_S(H)$ is the Sommerfeld coefficient derived from the magnetization data using the Maxwell relation,\cite{Hardy} and $c_i$ ($i=a, b$, and $c$) are field-independent arbitrary parameters.
This relation is based on an assumption that $1/T_1T$ is proportional to the square of the density of states at the Fermi level, or the square of the effective mass.
The $c_i$ actually depends on the hyperfine coupling constant and the anisotropy of the magnetic fluctuations.
The curves of $c_i \gamma_S^2(H)$ almost reproduce the field dependence of $1/T_1T$ for both fields, and indicate that $1/T_1T$ for $H_0 \parallel c$ gradually recovers toward $H_0 \rightarrow 0$, whereas $1/T_1T$ for $H_0 \parallel b$ approaches constant value in lower fields.
The extrapolation value at $H_0 \rightarrow 0$ is expected to be somewhat larger in $H_0 \parallel b$.
This difference in $1/T_1T$ corresponds to the difference between $\chi"_{b}(q,\omega_N)$ and $\chi"_{c}(q,\omega_N)$ from Eq.(2), because the hyperfine coupling constants are almost same between $H_0 \parallel b$ and $H_0 \parallel c$.

In UCoGe, the large difference of $\sim10$ times has been observed between $1/T_1T$ for $H_0 \parallel b$ and $H_0 \parallel c$, indicative of the strong Ising character of the magnetic fluctuations.\cite{Ihara,Hattori}
Such a large difference is not seen in URhGe.
We should pay attention to the fact that $1/T_1T$ for $H_0 \parallel b$ is so sensitive to the field along the $c$ axis induced by the misalignment in UCoGe.\cite{Hattori}
Actually, the coefficient $A$ in resistivity is suppressed drastically by the low magnetic field along the $c$ axis.\cite{Aoki_UCoGe}
Using the relation of $\gamma_S \propto \sqrt{A}$, the initial slope near zero field in the reduction of $\gamma_S$ is estimated to be $d[(\gamma_S(H)-\gamma_S(0))/\gamma_S(0)]/dH \sim -30$\%/T in UCoGe.
In URhGe, $d[(\gamma_S(H)-\gamma_S(0))/\gamma_S(0)]/dH \sim -7$\%/T is estimated for $H_0 \parallel c$.\cite{Hardy}
Thus, we consider that the sensitivity in $1/T_1T$ to the misalignment from the $b$ axis is not so strong in URhGe as UCoGe.
In URhGe, the small anisotropy in $1/T_1T$ at low fields and low temperatures suggests that the Ising character of the magnetic fluctuations well below $T_{\rm Curie}$ is not so strong as UCoGe, although a future confirmation using a single crystal is desired.
An importance of the longitudinal magnetic fluctuation for superconductivity has been suggested in UCoGe.\cite{Hattori}
The weaker Ising character of the magnetic fluctuations below $T_{\rm Curie}$ in URhGe is likely to be relevant to induce the difference in $T_{sc}$ between UCoGe ($T_{sc}=0.6$ K) and URhGe ($T_{sc}=0.25$ K).

Next, we move to a discussion for the magnetic fluctuation at around $H_R$.
For $H_0 \parallel b$, $1/T_1T$ increases toward $H_R$.
This $1/T_1T$ corresponds to the magnetic fluctuations along both the $a$ and $c$ axes, but it is considered that the contribution along the $c$ axis is dominant, because the susceptibility along the $a$ axis is small in URhGe.\cite{Hardy,Knafo}
On the other hand, we observed that $1/T_2$ also increases significantly toward $H_R$.
In general, $1/T_2$ is expressed as follows.
\begin{eqnarray}
\frac{1}{T_2} = \frac{1}{2T_1} + \lim_{\omega \rightarrow 0} \frac{\gamma_N^2}{2} \int_{-\infty}^{\infty} \langle \delta H^z(t) \delta H^{z}(0) \rangle \exp(-i \omega t) dt
\end{eqnarray}
Here $\langle \delta H^z(t) \delta H^z(0) \rangle$ are time correlation function for magnetic fluctuations parallel to the quantum axis of the nuclear spin.
Thus, $1/T_2$ corresponds to the magnetic fluctuations along the external magnetic field, that is, along the $b$ axis.
The large enhancement of $1/T_2$ toward $H_R$ indicates that the magnetic fluctuations along the $b$ axis are enhanced, in addition to the enhancement of the magnetic fluctuations along the $c$ axis detected by $1/T_1T$.
We cannot determine which of fluctuations along the $b$ the $c$ axes is stronger, because quantitative evaluation for $1/T_2$ is difficult.

As shown in Fig.~9, the temperature dependence of $1/T_1T$ shows a large increase toward low temperatures in the PM phase at 12 T, where the magnetic moment is expected to lie along the $b$ axis.
Thus, it is clear that the magnetic fluctuations perpendicular to the magnetic moment (transverse fluctuation) develop at around $H_R$.
This is in sharp contrast to the strong-Ising case in a metamagnetic transition of UCoAl, where a critical endpoint of the metamagnetic transition is induced at a finite temperature by a magnetic field along the $c$ axis (easy axis).
In UCoAl, the magnetic fluctuations are observed in $1/T_2$, whereas they are not detected in $1/T_1T$, when the magnetic field is applied along the $c$ axis, \cite{Nohara,Karube} 
The contrasting behavior between $1/T_1T$ and $1/T_2$ indicates that the magnetic fluctuation has strong Ising anisotropy in UCoAl.
In URhGe, magnetic fluctuations are observed both in $1/T_2$ and $1/T_1T$.
This indicates that the magnetic fluctuations at around $H_R$ include both components along the $b$ and $c$ axes, and are not strong Ising-type.
Theoretically, the two SC phases in URhGe have been reproduced by considering the coupling between electrons and softened magnon with transverse fluctuations.\cite{KHattori}
It is important to determine which of fluctuations along the $b$ and $c$ axes is crucial for field-induced superconductivity in URhGe.

\section{Conclusion}
We have performed $^{73}$Ge-NMR and NQR measurements for URhGe.
Two oriented samples made us to investigate the static and dynamical magnetic properties of URhGe for $H_0 \parallel c$ (easy axis) and $H_0 \parallel b$.
The observation of the phase separation for $H_0 \parallel b$ indicates that the spin reorientation is of first order at low temperatures.
The tricritical point was roughly estimated to be present between (10 T, 7 K) and (12 T, 4 K) on the field-temperature phase diagram for $H_0 \parallel b$.
The nuclear spin relaxation rate $1/T_1$ shows that the magnetic fluctuations were suppressed for $H_0 \parallel c$, whereas the fluctuations remain strongly under the magnetic field along the $b$ axis.
The extrapolation values toward $H_0 \rightarrow 0$ are likely to suggest that the Ising character of the magnetic fluctuations well below $T_{\rm Curie}$ in URhGe is not so strong as UCoGe.
Toward the spin reorientation filed for $H_0 \parallel b$, the enhancements of $1/T_1T$ and the nuclear spin-spin relaxation rate $1/T_2$ were observed.
These indicate that the magnetic fluctuations along the $b$ axis and the $c$ axis develop at around the first-order spin reorientation, which are the driving force of the field-induced superconductivity in URhGe.

\section*{Acknowledgement}

We acknowledge H. Mukuda for experimental support, and D. Aoki, I. Sheikin, and K. Hattori for valuable discussions.
This work has been partly supported by Grants-in-Aid for Scientific Research (Nos. 24340085 and 26400359) from the Ministry of Education, Culture, Sports, Science, and Technology (MEXT) of Japan.


\begin{thebibliography}{99}
\bibitem{Vzz}
In the calculation, $V_{zz}$ was estimated to be tilted by 24$^{\circ}$ from the $a$ axis to the $c$ axis, but we did not use this angle in the analysis of the spectrum. The calculation also provides $\nu_Q^{cal}=1.32$ MHz and $\eta^{cal}=0.24$.
\end{thebibliography}
\end{document}